\documentclass[conference]{IEEEtran}
\IEEEoverridecommandlockouts
\usepackage{etoolbox}
\makeatletter
\patchcmd{\@makecaption}
  {\scshape}
  {}
  {}
  {}
\makeatother
\usepackage{cite}
\usepackage{color}
\usepackage{xcolor,colortbl}
\definecolor{Gray}{gray}{0.85}
\usepackage{amsmath,amssymb,amsfonts}
\usepackage{algorithmic}
\usepackage{graphicx}
\usepackage{textcomp}
\usepackage{xcolor}
\def\BibTeX{{\rm B\kern-.05em{\sc i\kern-.025em b}\kern-.08em
    T\kern-.1667em\lower.7ex\hbox{E}\kern-.125emX}}

\begin{document}

\title{Compact and Low-Loss PCM-based Silicon Photonic MZIs for Photonic Neural Networks}

\author{Amin Shafiee$^1$, Sanmitra Banerjee$^2$, Benoit Charbonnier$^3$, Sudeep Pasricha$^1$, and Mahdi Nikdast$^1$ \\
$^1$Department of Electrical and Computer Engineering, Colorado State University, Fort Collins, CO 80523, USA\\
$^2$NVIDIA, Santa Clara, CA 95134, USA \\
$^3$Université Grenoble Alpes, CEA, LETI, F38000 Grenoble, France}

\maketitle

\begin{abstract}

We present an optimized Mach--Zehnder Interferometer (MZI) with phase change materials for photonic neural networks (PNNs). With 0.2~dB loss, $-$38~dB crosstalk, and length of 52~$\mu$m, the designed MZI significantly improves the scalability and accuracy of PNNs under loss and crosstalk. 

\end{abstract}

\section{Introduction}
Silicon photonic (SiPh) integrated circuits have a wide range of applications, from sensing and high-speed chip-scale interconnects to energy-efficient optical computation in contemporary photonic computing and communication systems. 
 SiPh neural network accelerators based on a mesh of Mach--Zehnder Interferometer (MZI) devices  are being developed \cite{nature_ONN} to overcome the shortcomings of conventional electronic accelerators, in terms of energy efficiency and latency for the post-Moore era. Despite their promising performance, MZI-based photonic neural networks (PNNs) lack scalability due to inevitable fabrication-process variations (FPVs), large footprint of MZIs, and intrinsic optical losses and crosstalk noise in the underlying MZI devices \cite{shafiee2022loci,Tnano}. 
 
In this paper, leveraging an optimized multimode-interference-based beam splitter and combiner (MMI-BS) integrated with a compact and lossless phase-change material (PCM)-based phase shifter (PhS), we propose a compact, low-loss, and FPV-tolerant SiPh PCM-based MZI multiplier unit for PNNs.
The system-level results from simulating multiple PNNs show that using the optimized PCM-based MZI leads to an average 2.5\% inferencing accuracy loss, which is 71.5\% better compared to the PNNs using conventional MZIs.

\section{PCM-based MZI Design and Optimization}
\begin{figure}[t]
    \centering
    \includegraphics[width=0.48\textwidth]{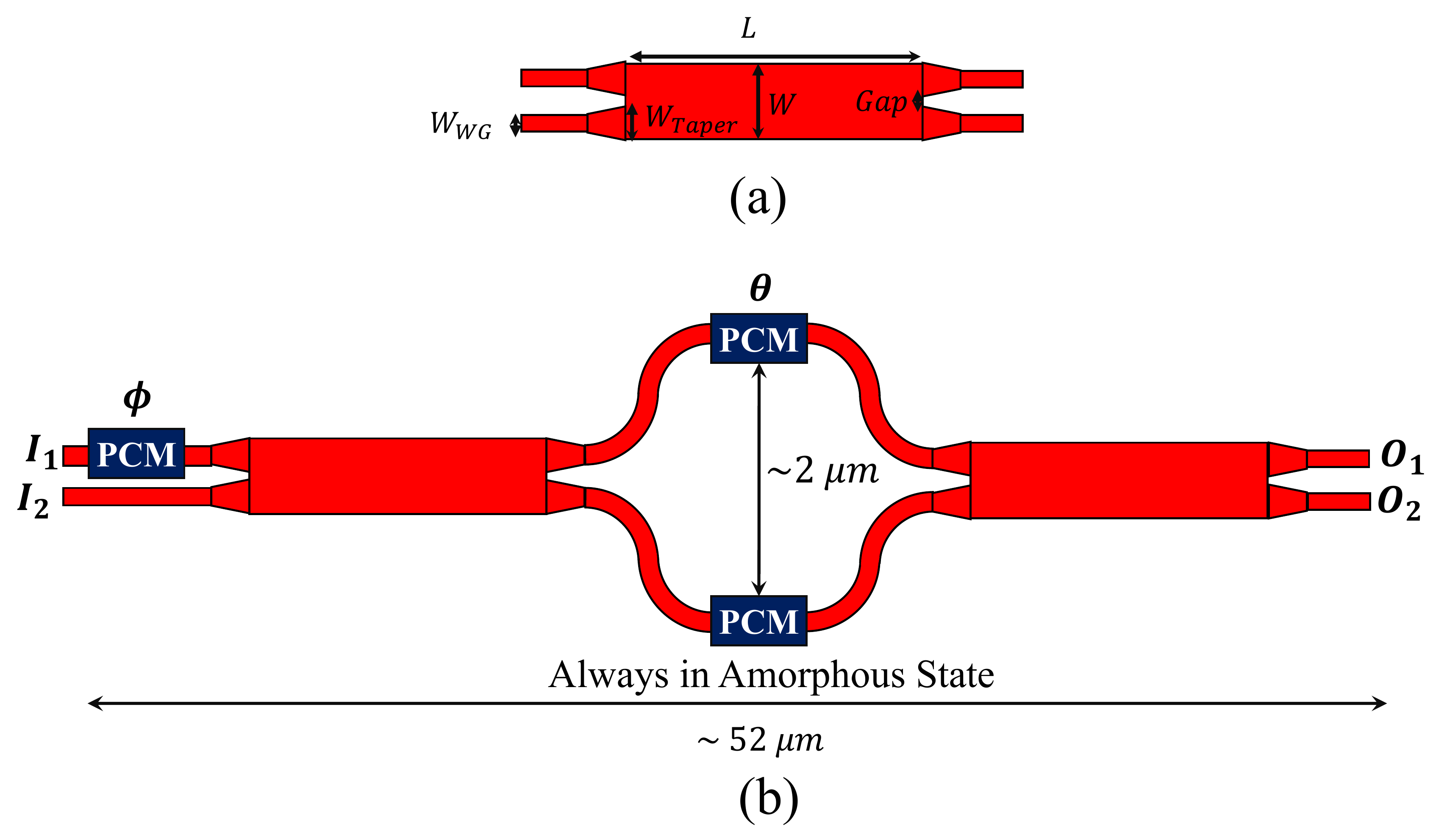}
    \caption{(a) Optimized MMI-based compact and lossless beam splitter. (b) Structure of the full MZI multiplier.}
    \label{MZI}
\end{figure}

 Compared to conventional directional couplers, MMI devices tend to offer lower crosstalk and higher reliability in achieving a 50:50 splitting ratio \cite{mahdian_MMI}. The structure of the designed SiPh 50:50 MMI-BS in this work is shown in Fig.~\ref{MZI}(a). A multi-variable simplex optimization method integrated with 3D electromagnetic simulations (finite-difference-time-domain (FDTD)) was used to optimize the geometry of the MMI-BS to minimize the splitting-ratio deviation from the ideal 50:50 splitting as well as the loss. The design-space parameters of the MMI-BS are reported in Table \ref{table1}.

\begin{figure}[t]
    \centering\includegraphics[width=0.48\textwidth]{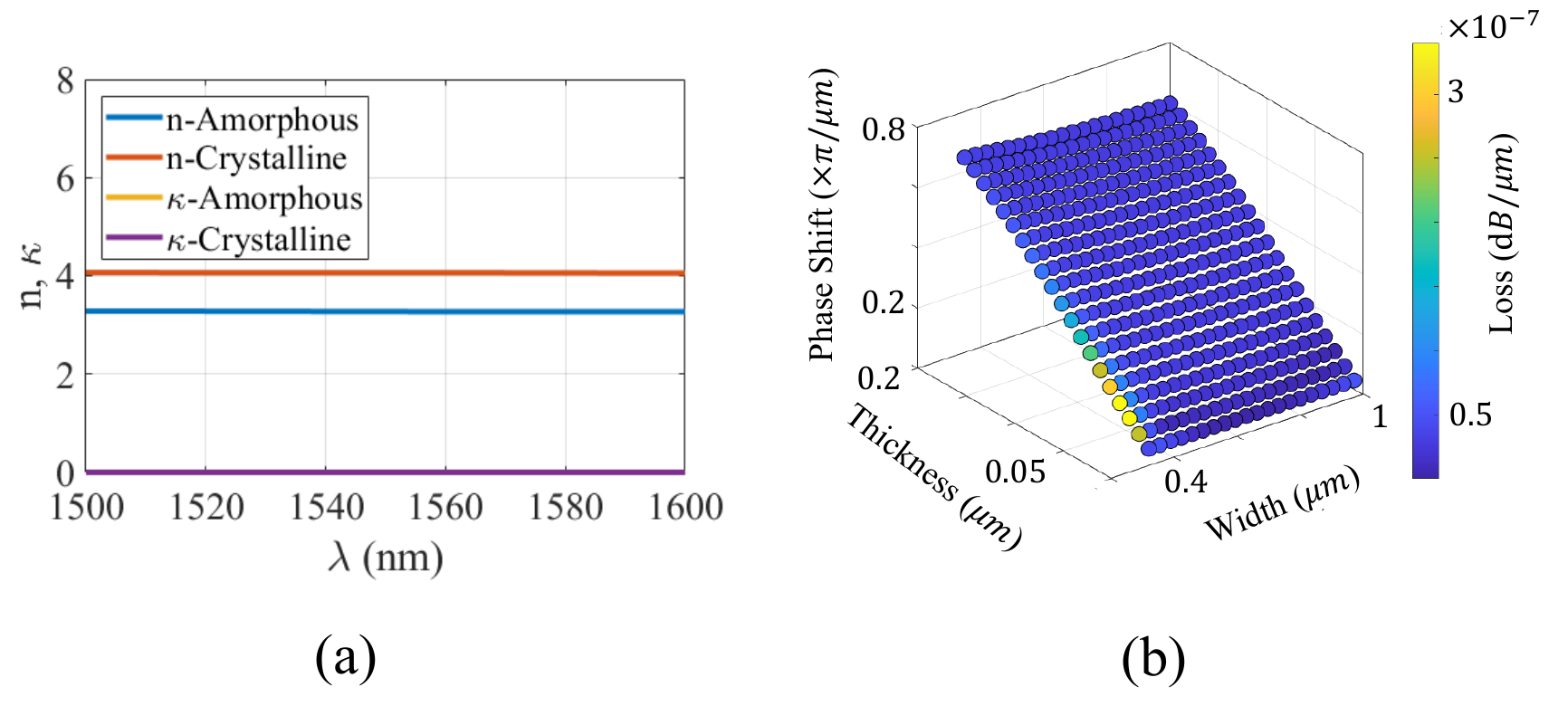}
    \caption{(a) Optical Properties of Sb$_2$Se$_3$ (b) Design-space exploration of Sb$_2$Se$_3$-based PhS.}
    \label{PS_design}
\end{figure}

\begin{figure*}[t]
    \centering
    \includegraphics[width=0.99\textwidth]{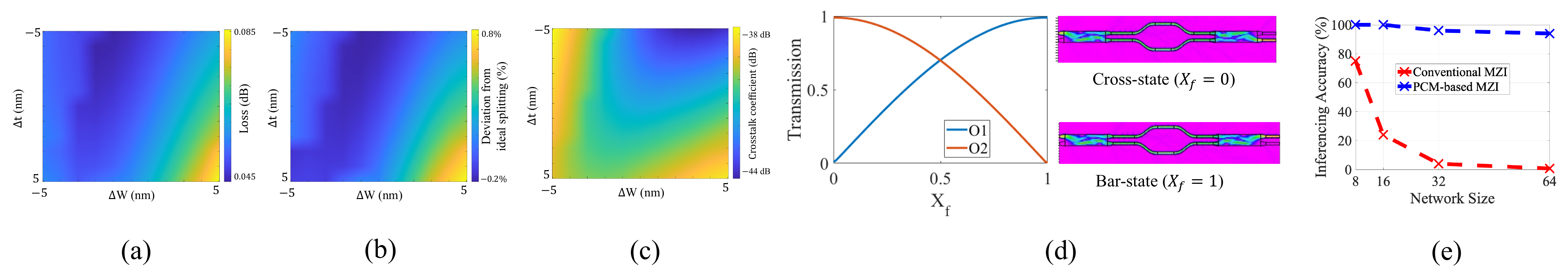}
    \caption{MMI-BS  loss, (a), and splitting-ratio deviation from 50:50, (b), under FPVs. (c) MZI's crosstalk coefficient under FPVs. (d) MZI's transmission using the designed PhS. (e) Inferencing accuracy of the PNNs of different sizes with the Clements configuration using the PCM-based and conventional MZIs. $\Delta t$: SOI thickness variation, $\Delta W$: Width variation, $X_f$: Crystalline fraction of the PCM which determines the portion of the material with crystalline state.}
    \label{fig2}
\end{figure*}

\begin{table*}[t]
    \centering
    \caption{Design parameters of the optimized MMI-BS (see Fig.~\ref{MZI}(a)).}
    \label{table_1}
\begin{tabular}{|c|c|c|c|c|c|} 
 \hline
 \rowcolor{Gray}
  $L$ ($\mu$m) & $W$~($\mu$m) & Gap~($\mu$m) & $W_{Taper}$~($\mu$m) & $W_{WG}$~($\mu$m) & SOI Thickness ($\mu$m) \\ 
 \hline
 15.7312 & 2.08320  & 0.40208 & 0.90532& 0.5& 0.22 \\ 
 \hline
\end{tabular} \label{table1}
\end{table*}

A PhS is an integral part of MZI-based multipliers in PNNs which are responsible for performing matrix-vector multiplication by determining the amount of light that moves from one arm to another by inducing $\theta$ and $\phi$ phase shifts between the MZI's arms (see Fig.~\ref{MZI}(b)). To design the required PhS in the MZI multiplier, we opted to use PCMs. PCMs transform from amorphous to crystalline state, and vice versa, when heated by an external heat source. This results in different optical and electrical properties. The optical properties of Sb$_2$Se$_3$ in the amorphous and crystalline state using the Lorentz model 
\cite{shafiee_survey} is depicted in Fig.~\ref{PS_design}(a)~\cite{shafiee2023design}. We can see that the extinction c	oefficient of the material is zero in the C-band (1530--1565~nm), which makes Sb$_2$Se$_3$ lossless. When deposited on top of a silicon-on-insulator (SOI) waveguide, the contrast between the optical properties in different states leads to a change in the effective index of the underlying waveguide \cite{shafiee2023design,shafiee_survey}. The difference in the effective index of the waveguide between the crystalline and amorphous state as well as PCM's lossless behaviour makes Sb$_2$Se$_3$ an excellent candidate to implement a compact and lossless PCM-based PhS \cite{rios2022ultra}. 

We performed a design-space exploration for the Sb$_2$Se$_3$-based PhS in Fig.~\ref{PS_design}(b). Observe that for all of the design points considered, the PhS exhibited a negligible loss. Taking into account a waveguide width of 500~nm for the designed BS-MMI, the thickness of 70~nm was selected for Sb$_2$Se$_3$ on top of the SOI waveguide to ensure maximum optical transmission and phase shift per unit length. This design achieves a phase shift per length of 0.2~$\pi$/$\mu$m which leads to a compact $\approx$5~$\mu$m PhS to induce a $\pi$ phase shift when the PCM is fully crystallized. Compared to \cite{baghdadi2021dual_NOEM}, the proposed PhS offers lossless performance and zero static power consumption to maintain the phase shifts with $\approx$40\% shorter length.

Leveraging the optimized MMI-BS and PCM-based PhS, a full MZI is designed (see Fig.~\ref{MZI}(b)). Note that 
a microheater similar to the one presented in \cite{shafiee2023design} can be used here to trigger the phase change of the PCM. Moreover, a minimum distance of 2~$\mu$m was considered between the arms to avoid any thermal crosstalk between the PCMs when programming them \cite{shafiee_survey}. Including 8-$\mu$m S-bends in the MZI's arm, the total length of the MZI is $\approx$52~$\mu$m (area: $\approx$2$\times$52~$\mu$m$^2$). Note that to make the design balanced, the phase of the PCM on the lower arm of the MZI is always maintained at the amorphous state to ensure no phase difference between the arms when $\theta=$~0. 

\section{Results, Discussions, and Conclusion}
Prior work in \cite{shafiee2022loci, Tnano} showed the severe impact of optical loss, crosstalk noise, and FPVs on the scalability and accuracy of MZI-based PNNs. Optical loss and crosstalk noise accumulate as PNNs scale up, and hence deteriorate signal integrity in PNNs and limit their scalability and accuracy \cite{shafiee2022loci}. Moreover, FPVs cause phase noise and splitting-ratio deviations in the MZIs, which impact the network accuracy \cite{Tnano}. Here, we show how our optimized MZI can help alleviate these issues in PNNs. Using FDTD simulations, the designed MMI-BS shows a 50:49 splitting-ratio with 0.04 dB loss at 1550~nm when the nominal design is used (see Table \ref{table_1}). Moreover, the sensitivity of the MMI-BS's loss to $\pm$5~nm variations in both waveguide width and thickness is shown in Fig.~\ref{fig2}(a). Observe that the loss of the MMI-BS can be as high as 0.085~dB under FPVs. The work in \cite{shafiee2022loci} showed that for up to 0.22~dB loss in MZIs' directional couplers (DCs), the inferencing accuracy of a 16$\times$16 PNN with two hidden layers trained on the MNIST dataset drops by less than 5\%. Considering 0.22~dB as a loss threshold for DC's operation, the MMI-BS offers acceptable performance for PNNs under  FPVs.

The splitting-ratio deviation (from 50:50) of the MMI-BS under FPVs is depicted in Fig.~\ref{fig2}(b). Observe that under the considered FPV range, the MMI-BS exhibits a minimal splitting-ratio deviation of only $-$0.2\% to 0.8\%, showing high robustness under FPVs. 
Accordingly, the resulting MZI offers up to 0.2~dB loss and $-$38~dB crosstalk (see Fig.~\ref{fig2}(c)), which are substantially lower than MZI multipliers currently used in PNNs \cite{shokraneh_theoretical}. Fig.~\ref{fig2}(d) shows the output transmissions of the proposed MZI and the electric-field profile of the MZI in Bar- and Cross-state using electromagnetic simulations. Using only $I_1$ and terminating $I_2$ (see Fig.~\ref{MZI}(b)), observe that when the crystalline fraction ($X_f$) approaches 1, the phase shift reaches $\pi$, moving the MZI to the Bar-state.

To evaluate the system-level impact of the proposed PCM-based MZI when used in PNNs, four PNNs with input layer dimensions of 8$\times$8, 16$\times$16, 32$\times$32, and 64$\times$64 connected to two fully-connected layers (with the same dimensions as the input layer) using Clements configuration \cite{shafiee2022loci} were trained on a Gaussian dataset presented in\cite{shokraneh2020diamond}. The nominal accuracy was 100\%. As it can be seen from Fig. \ref{fig2}(e), the inferencing accuracy using conventional MZIs (using the same optical loss and crosstalk values in \cite{shafiee2022loci}) drops by 98\% for the 64$\times$64 PNN. Yet, the same PNN designed based on the proposed PCM-based MZI shows only up to 6\% accuracy drop. 
Note that even under FPVs in MMI-BSes, the PNNs implemented using our PCM-based MZIs show no additional inferencing accuracy loss (when the phase noise is neglected \cite{Tnano}). 

In summary, we designed a compact, low-loss, and FPV-tolerant PCM-based MZI multiplier that can help alleviate the severe impact of optical loss, crosstalk, and FPVs in MZI-based photonic systems such as PNNs and switching networks.
\section*{ACKNOWLEDGMENT}
This work was supported by the National Science Foundation (NSF) under grant number CNS-2046226 and CCF-1813370.




\bibliographystyle{IEEEtran}
\bibliography{IEEEabrv,main}

\end{document}